\def\plotone#1{\centering \leavevmode
\epsfxsize=\textwidth \epsfbox{#1}}

\documentstyle[11pt, aas2pp4]{article}    

\def\sqig{$\sim$}

\def\degrees{$^{\circ}$}

\def\source{2S0114+650}
\def\src{2S0114+650}
\def\LS{LS\ I\ +65\degrees010}

\begin{document}
\title{
Evidence for a very slow X-ray pulsar in \src\ from\\
RXTE All-Sky Monitor Observations}

\author{Robin H.D. Corbet\altaffilmark{1, 2}, John P. Finley\altaffilmark{3},
\and
Andrew G. Peele\altaffilmark{1, 4}}

\altaffiltext{1}{Code 662, NASA/Goddard Space Flight Center, Greenbelt, MD 20771}
\altaffiltext{2}{Universities Space Research Association;
corbet@lheamail.gsfc.nasa.gov}
\altaffiltext{3}{Department of Physics, Purdue University, 1396 Physics
Building, West Lafayette, IN 47097-1396;
finley@purds1.physics.purdue.edu}
\altaffiltext{4}{National Research Council Research Associate; peele@lheamail.gsfc.nasa.gov}

\begin{abstract}
Rossi X-ray Timing Explorer (RXTE) All-Sky Monitor (ASM) observations
of the X-ray binary \src\ show modulations at periods close to both the
optically derived orbital period (11.591 days) and proposed pulse
period (\sqig 2.7 hr). The pulse period shows frequency and intensity
variability during the more than 2 years of ASM observations analyzed.
The pulse properties are consistent with this arising from accretion
onto a rotating neutron star and this would be the slowest such period
known.  The shape of the orbital light curve shows modulation over the
course of the entire orbit and a comparison is made with the orbital
light curve of Vela X-1. However, the expected phase of eclipse, based
on an extrapolation of the optical ephemeris, does not correspond with
the observed orbital minimum. The orbital period derived from the ASM
light curve  is also slightly longer than the optical period.

\end{abstract}

\section{Introduction}

The X-ray source \src\ has an early spectral type supergiant optical
counterpart, \LS, with a recent classification of B1 Ia (Reig et al.
1996) and this object is thus thought to be a high mass X-ray binary.
With a distance of 7.2 kpc derived from this spectral classification
the X-ray luminosity is 1$\times$10$^{36}$ergs s$^{-1}$.  An orbital
period of 11.591$\pm$0.003 days was reported from optical radial
velocity measurements made by Crampton et al. (1985). However, the
optical Doppler curve did not allow a distinction to be made between a
circular orbit and one of small eccentricity (e = 0.16 +/- 0.07;
orbital period = 11.588 $\pm$ 0.003 days).  If the compact object in
\src\ is a neutron star then, in common with most other high mass X-ray
binaries, X-ray pulsations would be expected to be seen.  Although
there have been some reports of a pulsation period near 850 to 895
seconds (Yamauchi et al.  1990, Koenigsberger et al. 1983) the evidence
for a period of this length appears to be, at best, weak.  In contrast,
Finley, Belloni, \& Cassinelli (1992) reported a 2.78 $\pm$ 0.01 hr
period for which there is evidence from observations with several
satellites (ROSAT, Ginga, and EXOSAT) although the number of 2.78 hr
cycles in each observation was limited. The same period was also
apparent in further ROSAT observations (Finley, Taylor, \& Belloni
1994) and a lower limit to the Q-value ($P/\Delta P$) of 17 was derived.
If this is indeed the pulse period of a neutron star then it
would be the longest so far detected. The other known X-ray pulsators
have periods between 69ms and 1400 seconds (e.g. Bildsten et al.
1997).  However, alternative interpretations of such a modulation may
be possible such as instabilities in the primary star.  Some earlier
optical observations of \src\ gave somewhat different spectral
classifications and van Kerkwijk \& Waters (1989) have suggested that
the optical counterpart of \source\ displays some of the
characteristics of a Be star and some of a supergiant (see also
Guarnieri et al. 1994).  For a main-sequence luminosity classification
the distance to \src\ would be \sqig2.5 kpc and the luminosity
correspondingly less (cf. Reig et al. 1996).

Here we report on the results of an analysis of data obtained with the
All Sky Monitor (ASM) on board the Rossi X-ray Timing Explorer (RXTE)
covering a period of more than 2 years
which shows modulation on both the orbital period
and \sqig2.7 hr period.

\section{Data and Analysis}

The All-Sky Monitor detector on board RXTE (Bradt, Rothschild, \&
Swank 1983) is described in detail by Levine et al. (1996). The ASM
consists of three similar Scanning Shadow Cameras, sensitive to X-rays
in an energy band of approximately 2-12 keV, which perform sets of 90
second pointed observations (``dwells'') so as to cover \sqig80\% of
the sky every \sqig90 minutes.  The analysis presented here makes use
of both daily averaged light curves and light curves from individual
dwell data.  Light curves are available in three energy bands: 1.3 to
3.0 keV, 3.0 to 4.8 keV, and 4.8 to 12.2 keV. ASM observations of blank
field regions away from the Galactic center suggest that background
subtraction may yield a systematic uncertainty of about 0.1 counts/s
(Remillard \& Levine 1997).

RXTE ASM observations of \src\ have been obtained since January 1996,
with a short interruption due to instrumental problems in early 1996,
and the ASM light curve of \src\ is plotted in Figure 1. Because of the
low count rate from this source we show a rebinned and smoothed version
of the daily averaged light curve.  The data discussed in this paper
thus covers the period between MJD 50087 (1996 January 5) to MJD 50997
(1998 July 3).  The mean intensity of \src\ during this period is found
to be 0.35 counts/s or \sqig4.7 mCrab.

In order to search for periodic modulation in the ASM observations we
calculated a Fourier Transform of the light curve obtained from
individual dwell data weighting results by the errors on individual
data points. We also investigated the ``Lomb-Scargle'' (Lomb 1976,
Scargle, 1982) variation of the Fourier Transform and both techniques
gave comparable results.  The resulting power spectrum from the
weighted Fourier Transform technique is shown in Figure 2.  The two
strongest peaks in the power spectrum are located close to the 11.59
day orbital period of the system and the \sqig 2.7 hr period reported
by Finley et al. (1992). In addition, there is modest evidence for
power at peaks corresponding to harmonics of the orbital period. A peak
is also present at \sqig 96 minutes which corresponds to the orbital
period of RXTE. There is also a further peak of comparable strength at
approximately 25.65 hours which we suspect may be related to daily
variations in background levels - we find variability at
similar periods in at least some of the other sources that
have been observed with the ASM.
In Figure 3 we show a detail of the power spectrum near
the proposed orbital period.  We note that the strongest peak in the
power spectrum is apparently slightly offset from the period reported
by Crampton et al.  A sine wave fit to the RXTE ASM light curve yields
a period of 11.630 $\pm$ 0.007 days.

In Figure 4 we show a detail of the power spectrum near the Finley
et al. 2.7 hr period. In the figure it can be seen that the
strongest peak is offset from the value of the 2.78$\pm$0.01 hr found
by Finley et al. (1992). In addition, rather than a single peak, there
are multiple peaks suggesting period changes. To investigate period
changes in more detail we split the  ASM light curve into five sections
of equal length and computed individual power spectra. These are
shown in Figure 5 and demonstrate that the pulsations can exhibit
significant changes in both pulse strength and period.

The average pulse profile derived from the ASM modulations is shown
in Figure 6, this was constructed by dividing the light curve into
several sections and independently determining the period and phase
of the modulation during that section of data before co-adding. The mean
pulse profile appears to be approximately sinusoidal.

We also searched for variability on the proposed 850 second timescale
in the ASM data. In the range 800 to 900 seconds we find no evidence
for any periodic modulation with an amplitude greater than 0.062
cts/s.  In contrast, ASM observations of the 837 second pulsator X Per, which
has a mean flux of 0.66 cts/s (\sqig 9 mCrab) in the ASM, do show a
highly significant modulation with an amplitude of 0.13 cts/s
demonstrating that, for a sufficiently large amplitude, a period of
this length could have been detected with the ASM.

\section{Discussion}

\subsection{Orbital Modulation}
In Figure 7 we show the RXTE ASM data folded on the period obtained by
Crampton et al. (1985). By comparison with other high-mass X-ray
binaries, two ways in which we might expect to see orbital modulation
would be either (i) flares due to enhanced accretion at periastron
passage if the orbit is significantly eccentric, or (ii) an eclipse if
the orbital inclination is sufficiently high.  Marked on Figure 7 are
the phases during which either periastron passage or an eclipse would
be predicted based on the ephemerides reported by Crampton et al.  We
note that neither does the predicted periastron passage coincide with
the maximum of the folded light curve nor does the predicted eclipse
coincide with the minimum of the folded light curve.  If we require the
optical ephemeris to agree with the minimum of the ASM light curve we
would require the optical period to be somewhat longer than that
proposed by Crampton et al.  at 11.597 days.

As the power spectrum of the ASM light curve also suggests a
slightly different orbital period from that reported by Crampton et
al., we show in Figure 8 the ASM light curve folded on our best
period.  We note that this shows somewhat more of a ``saw-tooth''
profile than when folded on the optical period. We have examined the
folded light curves of other high-mass X-ray binaries observed with the
RXTE ASM and find a somewhat similar average behavior for Vela X-1, the
folded light curve of which is shown in Figure 9. (See also the ASCA
light curve from Feldmeier et al. 1996) In the case of Vela X-1 there
appear to be two components to the orbital variability: an eclipse plus
a modulation which persists over the entire orbit with the X-ray flux
peaking shortly after eclipse egress.  Vela X-1 has a similar orbital
period to that of \src\ at 8.96 days and an eccentricity of 0.088
(Bildsten et al.  1997).  The similarity of the light curves and
orbital periods suggest that a common mechanism may be
producing at least part of the modulation in both systems.  In order to
produce a modulation over the entire orbit a relatively large structure
is required. This could take the form of a gas stream such as an
accretion or photo-ionization wake (see e.g. Kaper,
Hammerschlag-Hensberge \& Zuiderwijk 1994).  The low count rate of
\src\ prevents a clear determination of whether an eclipse also occurs
in \src.  However, an RXTE PCA observation of \src\ by Finley et al.
(in preparation) suggests that the flux is very close to zero at this
orbital phase.  We have searched for any possible energy dependence in
the modulation of the light curve.  However, this is difficult because
of the relative hardness of the source - only the highest energy band
(4.8 - 12.2 keV) shows a very clear orbital modulation. In addition,
zero level offsets become even more important when the flux from this
relatively faint source is split into several bands. We thus cannot
draw any strong conclusions on spectral variability in \src\ from the
ASM data alone.

Given the failure of the optical ephemeris to match the folded X-ray
light curve we also fit an orbit to the velocities reported by Crampton
et al. and obtain essentially identical results.  The orbital periods
of \src\ derived from the optical and X-ray measurements are thus
apparently inconsistent. The difference of 0.04 $\pm$ 0.01 days appears
to be too large to be caused by a real change in the orbital period.
This would correspond to an increase in the orbital period at a rate of
\sqig3$\times$10$^{-5}$yr$^{-1}$, an order of magnitude faster, and also a
growth rather than a decay, compared to both theoretical predictions
and observations of other similar systems (cf. Brookshaw \& Tavani
1993, Rubin et al. 1997).
One possible explanation may be that the light curve and/or the radial
velocity curve are subject to variable distortions. For example,
X-ray irradiation could result in the variable contamination of
absorption lines by low equivalent width emission lines.  For
comparison, Smale \& Charles (1989) simulated such contamination in the
high-mass X-ray binary A0538-66 and were able to
produce apparent radial velocity shifts. Crampton et al. (1985) also note
possible variations in the velocity difference between measurements of
the H$\beta$ and H$\gamma$ absorption lines.  Alternatively, the cause
of the discrepancy may be variability in the X-ray orbital modulation.
If the X-ray modulation is produced by a mechanism
that is not strictly locked to orbital phase, for example absorption
in a gas stream, variations in the phase of X-ray modulation
could produce an apparent shift from the true orbital period.
To investigate changes in the orbital modulation of X-rays we
fitted sine waves to the first and second halves of the ASM light
curve separately. We obtained periods of 11.67 $\pm$ 0.02 d and
11.62 $\pm$ 0.03 d respectively. In addition, the amplitude of the
modulation declined from 0.17 $\pm$ 0.02 cts/s to 0.11 $\pm$ 0.02 cts/s.


\subsection{Pulsations}

There are two principal models which have been proposed to explain the
2.7 hr period modulation in \src. These are:

{\em Beta Cephei companion} - $\beta$ Cephei stars lie along a narrow
instability strip which encompasses late O and early B spectral types
and sub giant through super giant luminosity class, have optical
photometric pulsation periods
of 3 to 6 hours and ``typical'' amplitudes of a few hundredths of a magnitude.
As some optical observations had suggested spectral types in
the $\beta$ Cephei range (e.g. B0.5 III, Crampton et al. 1985),
Finley et al. (1992) and Taylor et al. (1995) considered whether
the 2.7 hr period in \src\ might arise from such pulsations in
the primary star.
Difficulties with this model are the lack of persistent
optical photometric modulation on this period
(Bell, Hilditch, \& Pollacco, 1993; Taylor et al. 1995)
and a clear explanation
of how this variability of the primary gives rise to
a modulation of the X-ray flux. 

{\em Rotating neutron star} - the majority of high-mass X-ray binaries
contain neutron stars and the rotation of a magnetized neutron star
naturally leads to X-ray modulation. The peculiarity here would be the
exceptionally long period compared to the known pulsation periods which
range from
69ms to 1400s.  If a neutron star in \src\ is rotating at an
equilibrium period of 2.7 hr then a magnetic field of
\sqig2-3$\times$10$^{13}$G is implied. Although this field strength
would be very high, some radio pulsars do have fields in this range
(e.g.  Taylor, Manchester, \& Lyne 1993). However, if the neutron star is
rotating more slowly than its equilibrium period, as may well be the case
for most wind-accretion powered X-ray pulsars (e.g. Corbet 1986), then the magnetic field
could naturally be lower. Waters \& van Kerkwijk (1989)
proposed that the spin periods in these systems are actually determined in an
earlier evolutionary phase before the companion star becomes a
super-giant. The peculiarities noted in the optical spectrum of
\LS\ could perhaps be an indication of an unusual evolutionary
history for the system.
Also in support of the rotating neutron star interpretation we note
that the variability of the \sqig 2.7 hr period length appears consistent
with the variations seen in other wind-accreting high-mass X-ray
binaries (e.g. Nagase 1989).

We also briefly consider two other phenomena which could, in principle,
give
modulation on a time scale of a few hours:

{\em ``EXO 2030+375-like variations''} -
The 2.7 hr period is somewhat reminiscent of the 3.96 hr
periodicity seen on one occasion in the Be/neutron star system EXO
2030+375 by Parmar et al. (1989).  In the case of EXO 2030+375 it is
already known that the neutron star rotation period is less than this
value at 42 seconds.  We note that a power spectrum of the RXTE ASM
light curve of EXO 2030+375 does not show any modulation on the 3.96
hr timescale reported by Parmar et al. (1989) even
though the source was active with a mean flux
of 0.35 cts/s (\sqig 4.6 mCrab) and the orbital modulation
was clearly detected.  For
\source\ possible periodicities of \sqig895s (Einstein) and
\sqig850s (Ginga) were reported but these are not present in
either our observations or HEAO-1, OSO 8, or EXOSAT data (Finley et al.
1992 and references therein). In contrast the 2.7 hr period is clearly
a persistent property of the source although variable in strength and
period length.

{\em Accreting White Dwarf} -
As noted above, at 2.7 hours this would be the longest known neutron
star rotation period.  However, non-synchronously rotating accreting
white dwarfs are known to have, in some cases, relatively long rotation
periods (e.g. Patterson 1994). The primary difficulty with interpreting
the 2.7 hr period as a white dwarf rotation period is the implied
X-ray luminosity which is too large to be explained by accretion onto a
white dwarf. In addition, changes in the length of
the 2.7 hr period are also difficult to explain
with a white dwarf model due to the larger moment of inertia of a white
dwarf compared to a neutron star.

X-ray spectral data also appear to indicate some support for
interpreting the 2.7 hr period as a neutron star rotation period.
Yamauchi et al. (1990) obtained observations of \source\ with Ginga and
detected an iron line at \sqig6.4 keV. The equivalent width was 0.34
keV during ``low'' states and 0.07 keV during ``flares'' (i.e. 2.7 hr
pulse maxima).  Yamauchi et al. report that during the flares the
equivalent width of the line decreased while the number of photons in
the line remained constant. As 6.4 keV is the energy for a fluorescent
iron line, one interpretation would be that the intrinsic source
luminosity remains constant and the variability is caused by geometric
factors - for example the pulsar beam moves out of our line of sight
while continuing to cause fluorescence in the cool wind of the primary
star.  This iron line behavior is apparently also present in
observations made with ASCA (Ebisawa 1997).

While the presence of persistent 2.7 hr pulsations, together with
optical spectra, makes the system a ``double lined binary" it does not
appear feasible to use the pulsations to determine the binary
parameters. The projected size of the orbit of the neutron star, $a_X
$sin$ i$, is estimated to be \sqig 130 light seconds, using the optical
velocity amplitude of 17 km s$^{-1}$ of Crampton et al. 1985 and
assuming a mass ratio of 15. This size corresponds to only \sqig1\% of
the length of the pulse period.

\section{Conclusions}
The X-ray light curve of \src\ shows the presence of modulation on both
the proposed orbital and 2.7 hr pulse periods of this system, although with a
small difference in the value of the orbital period compared to that
previously proposed.  Additional optical radial velocity measurements
will be important to enable an exact comparison between the phasing of
the X-ray light curve and time of expected eclipse. The continued
presence of the 2.7 hr modulation suggests that it does indeed
represent the rotation period of a neutron star.

\acknowledgments
This paper made use of quick look data provided by the RXTE ASM team at
MIT and GSFC. We thank many colleagues in the RXTE team for useful
discussions.

\pagebreak
\noindent
{\large\bf Figure Captions}

\figcaption[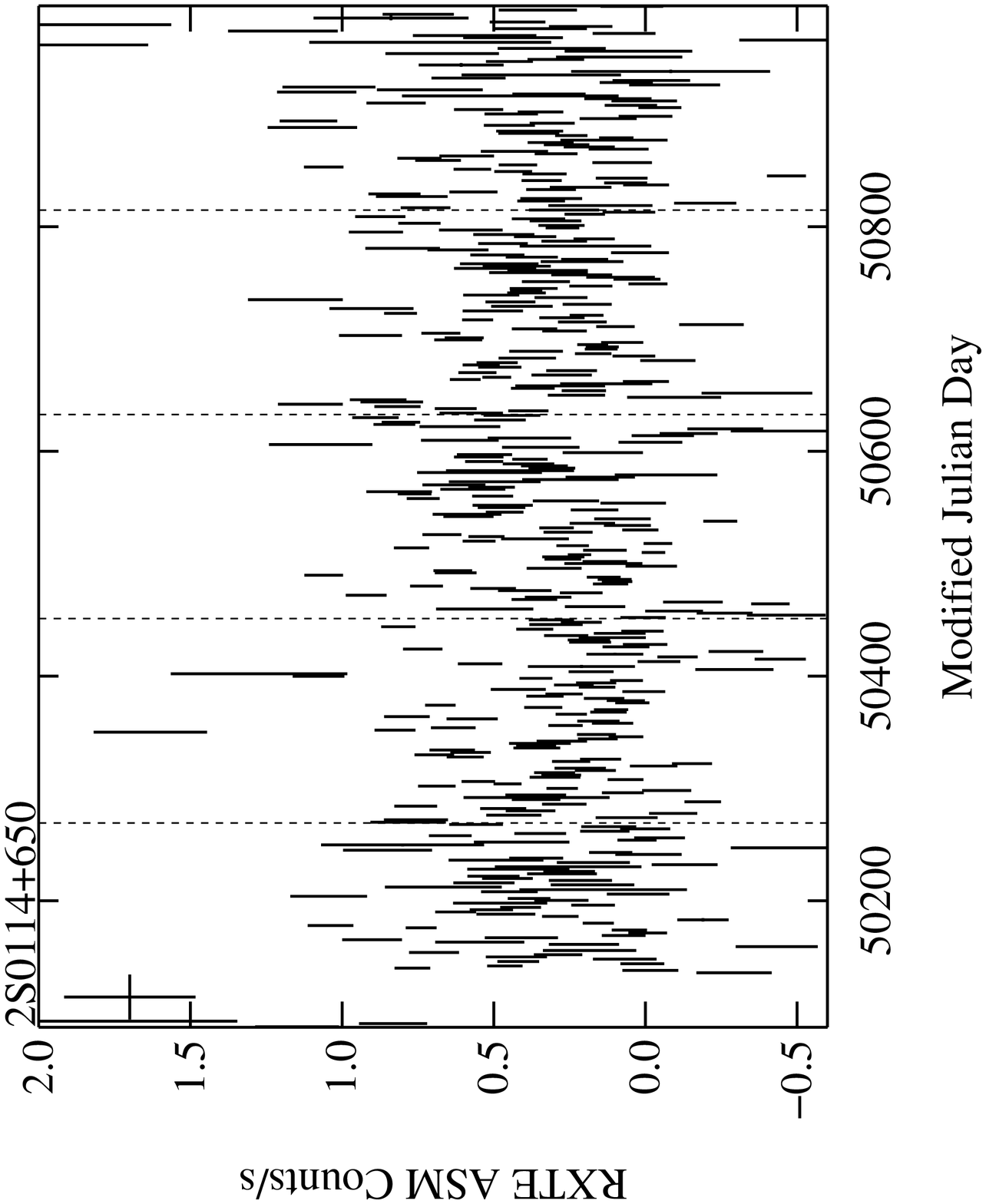]
{ASM light curve of \source. Data points are a rebinned and smoothed
version of the standard one day light curves. The dashed lines indicate
the sections the light curve was split into for computing the power
spectra plotted in Figure 5.}

\figcaption[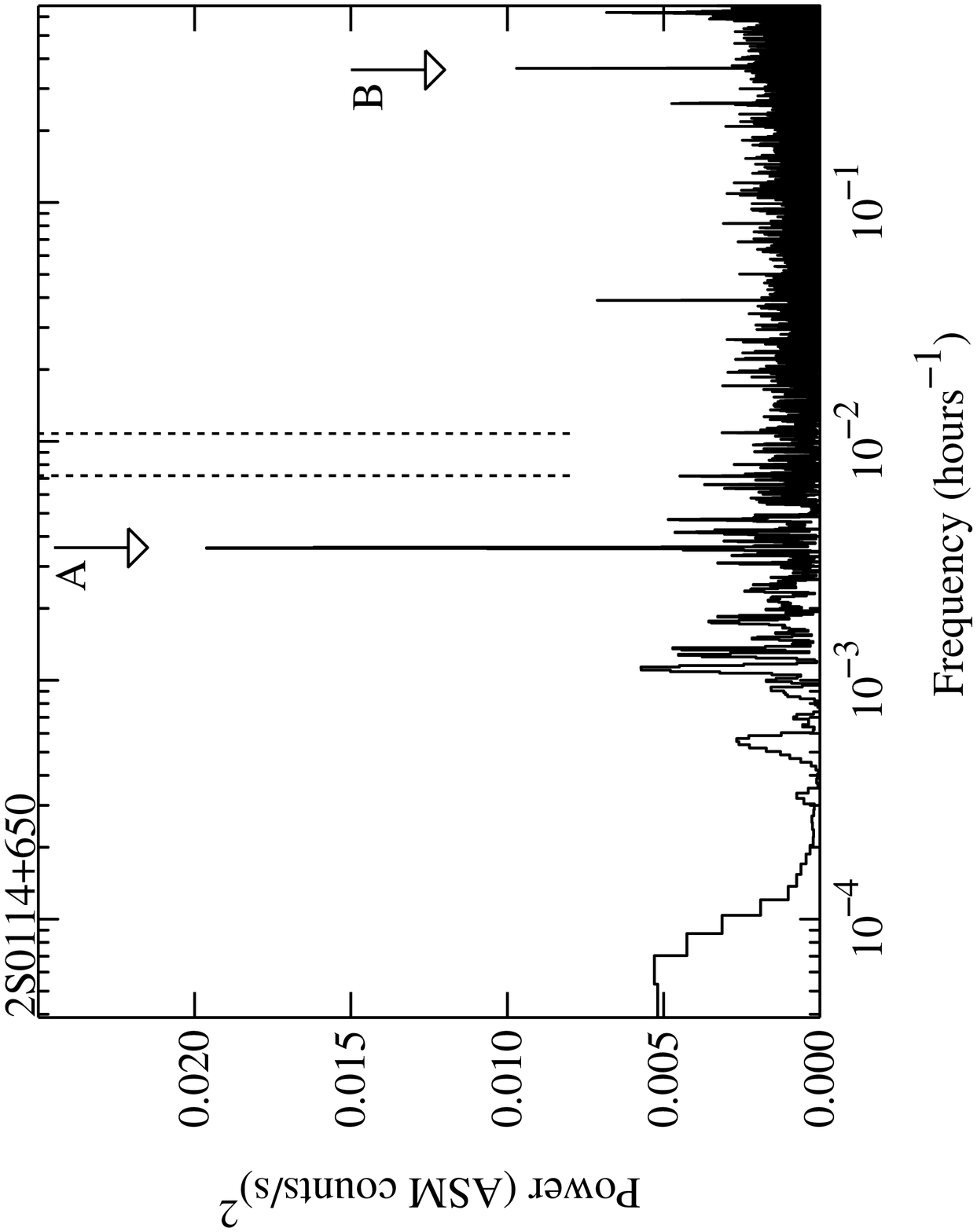]
{Power spectrum of the ASM light curve of \source. ``A'' and ``B''
indicate the periods reported by Crampton et al. (1985) and Finley et
al.  (1992) respectively. The dashed lines indicate harmonics of the
Crampton et al. (``A'') peak.}

\figcaption[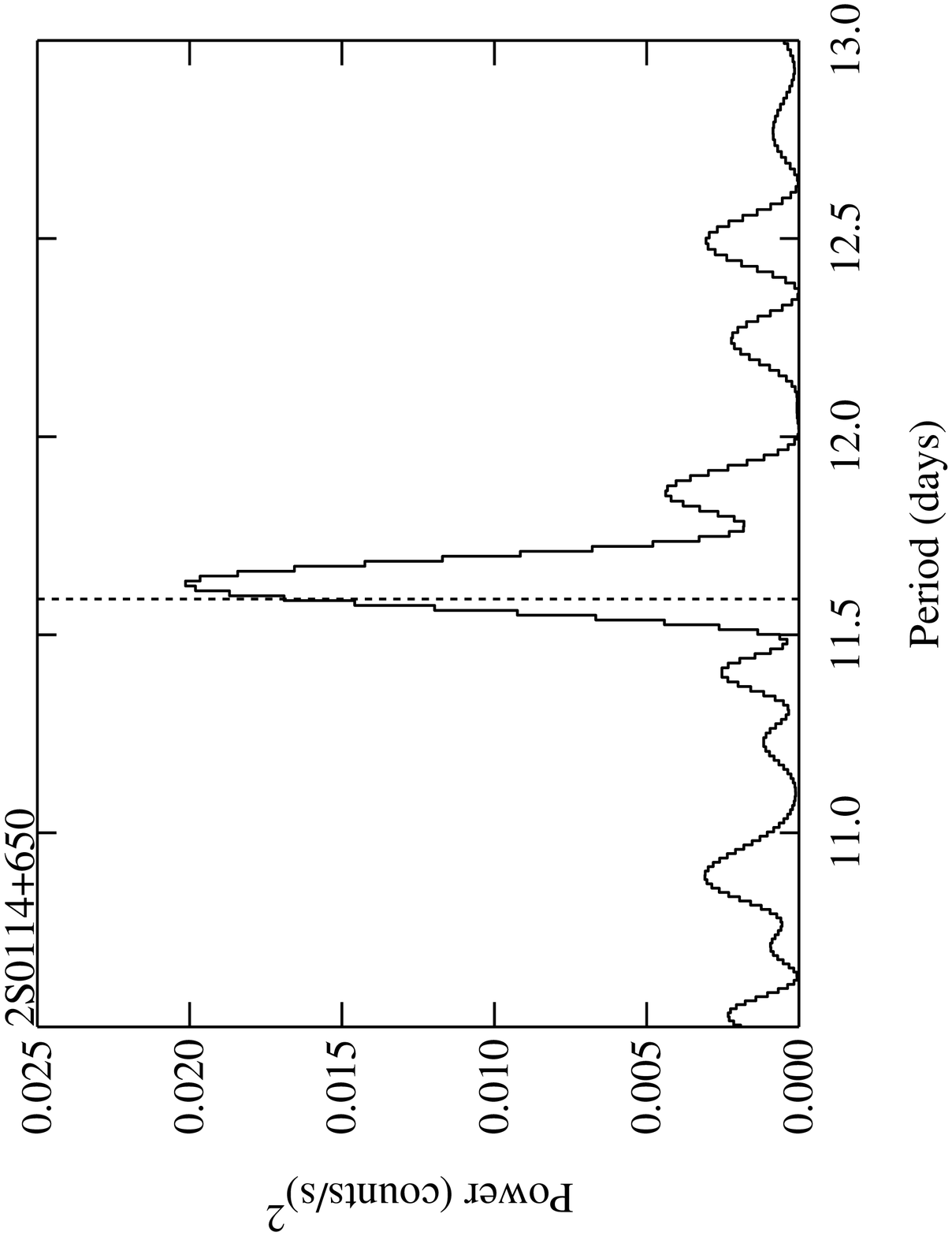]
{Close up of the peak in the power spectrum close to the Crampton et
al. (1985) optical period which is indicated by the dashed line.}

\figcaption[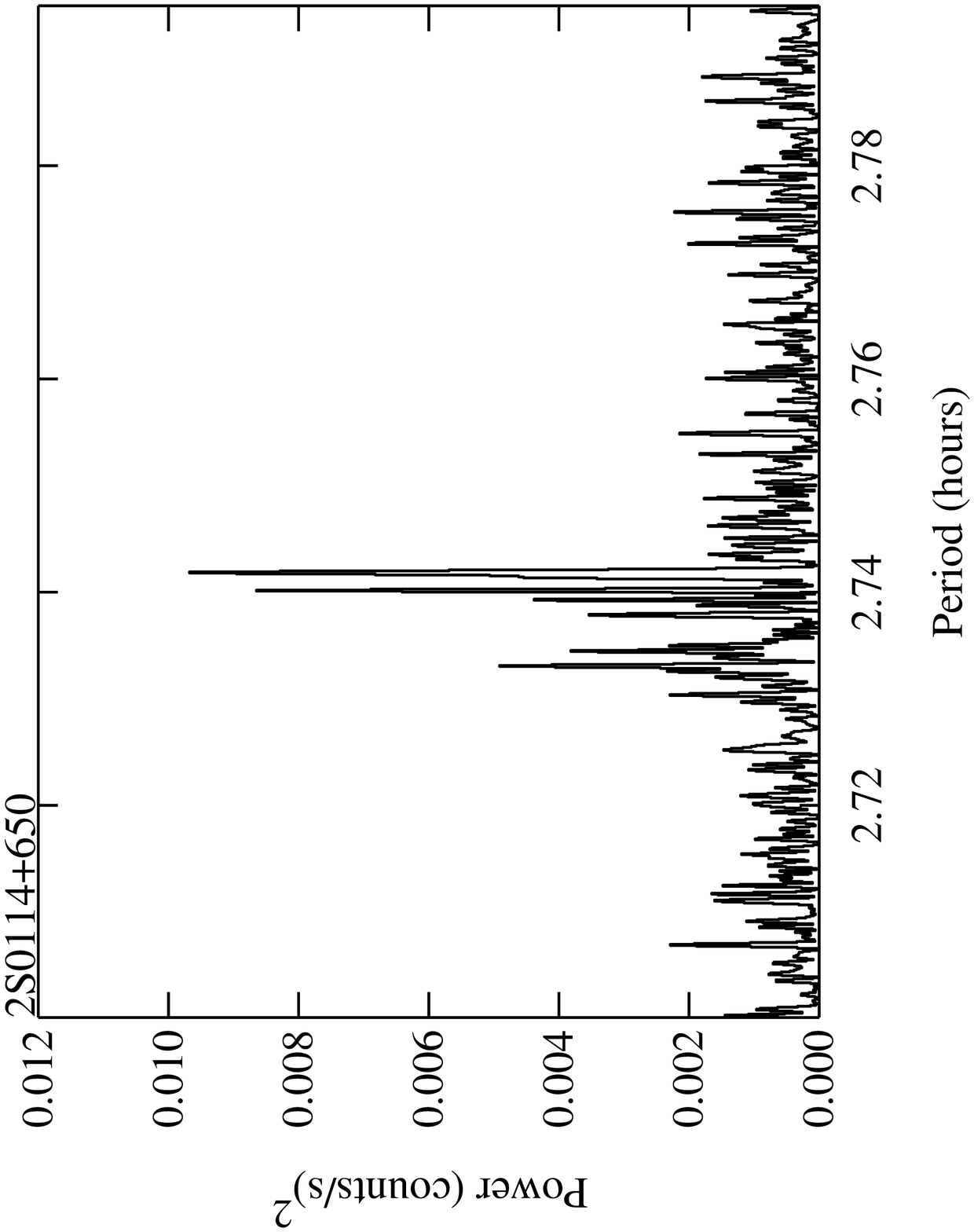]
{Detail of the power spectrum of the ASM light curve in the region
around the period reported by Finley et al. (1992).}

\figcaption[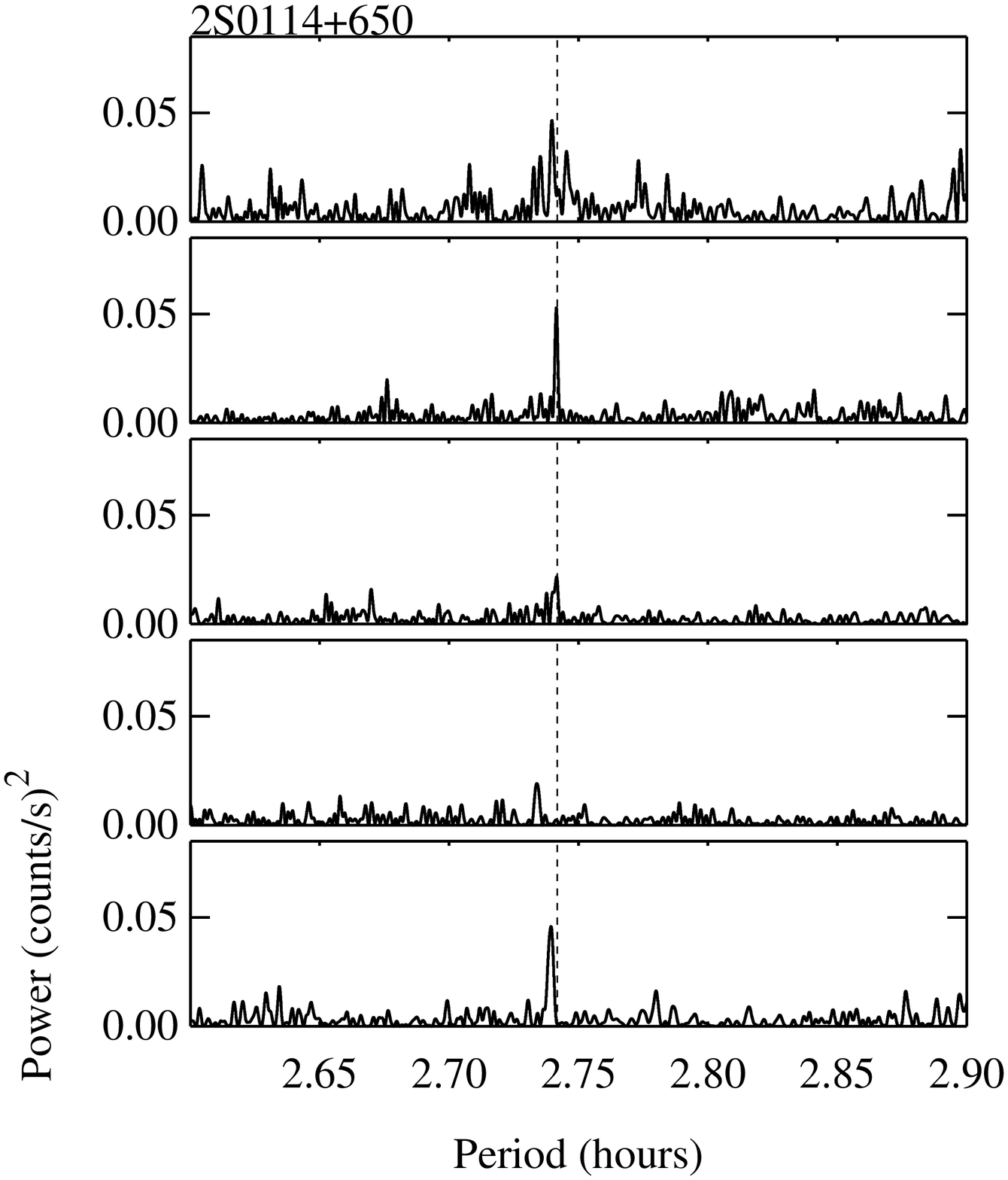]
{Power spectra of the ASM light curve around the Finley et al.
(1992) period. The light curve was divided into five sections of equal
length and power spectra calculated of each section individually.
The dashed lines indicate the period of the strongest peak in a power
spectrum of the entire data set. Time increases from the bottom to the
top of the figure and the intervals corresponding to the five data
sections are indicated in Figure 1 (MJD:  50087.4, 50269.2, 50451.0,
50632.9, 50814.7, 50996.6)}

\figcaption[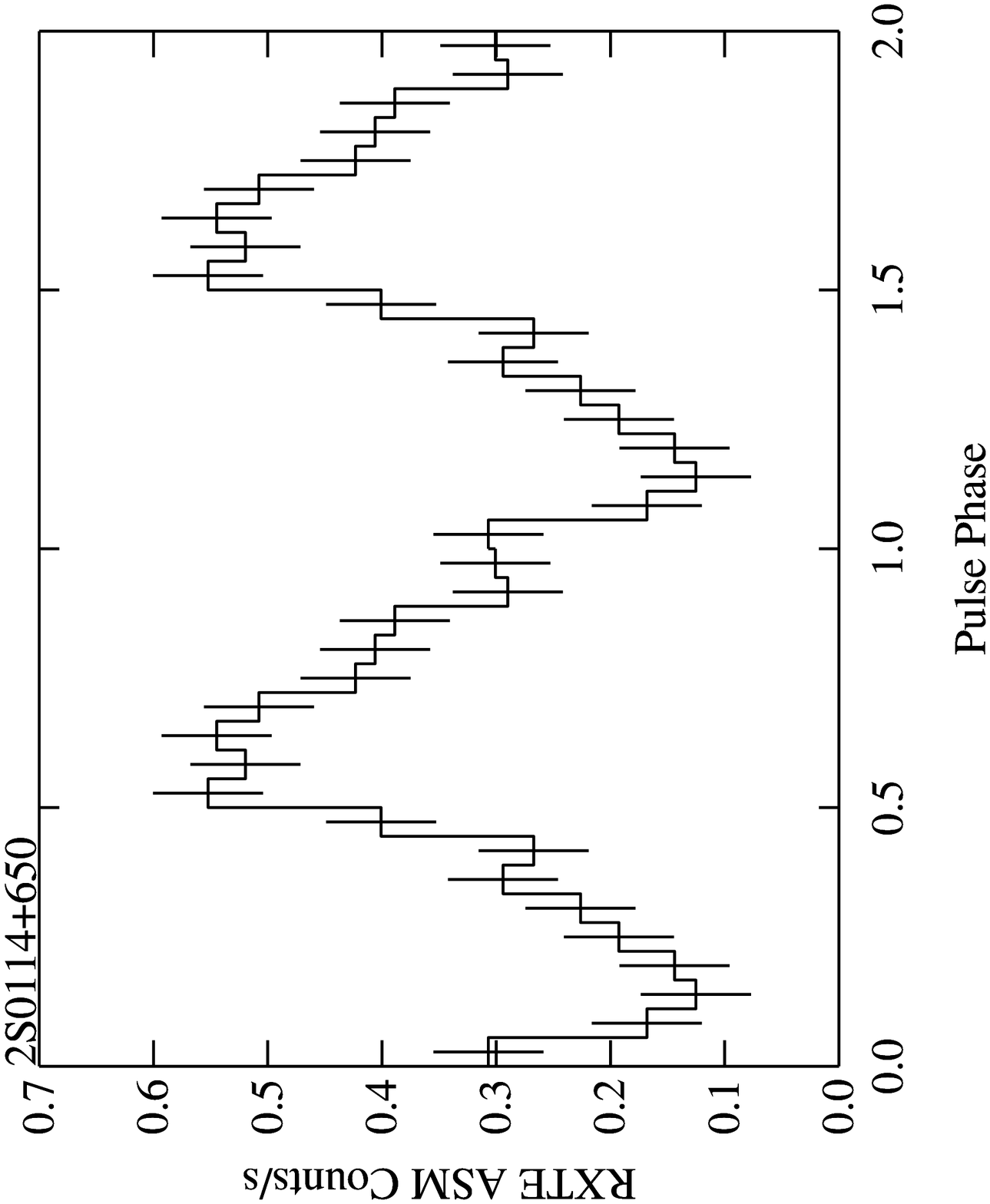]
{Mean \sqig 2.7 hr pulse profile of \source. Each data section identified
in Figure 1
was individually folded and phase shifted to form this mean profile.
Two cycles are shown for clarity.
}

\figcaption[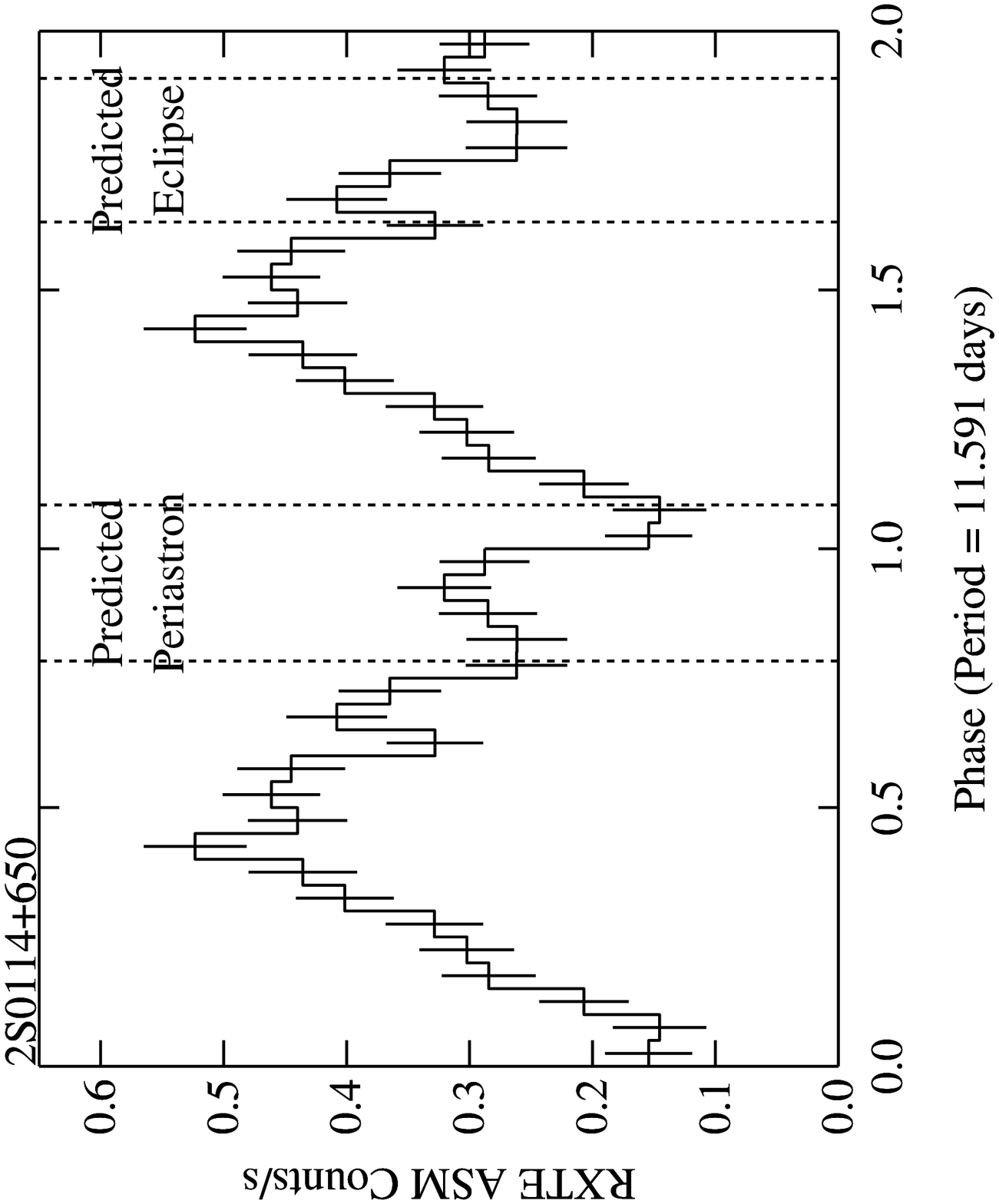]
{ASM light curve folded and binned on the Crampton et al. (1995)
period. Times of predicted eclipse and periastron passage based on the
Crampton et al. ephemeris are indicated. Two cycles are plotted for
clarity.}

\figcaption[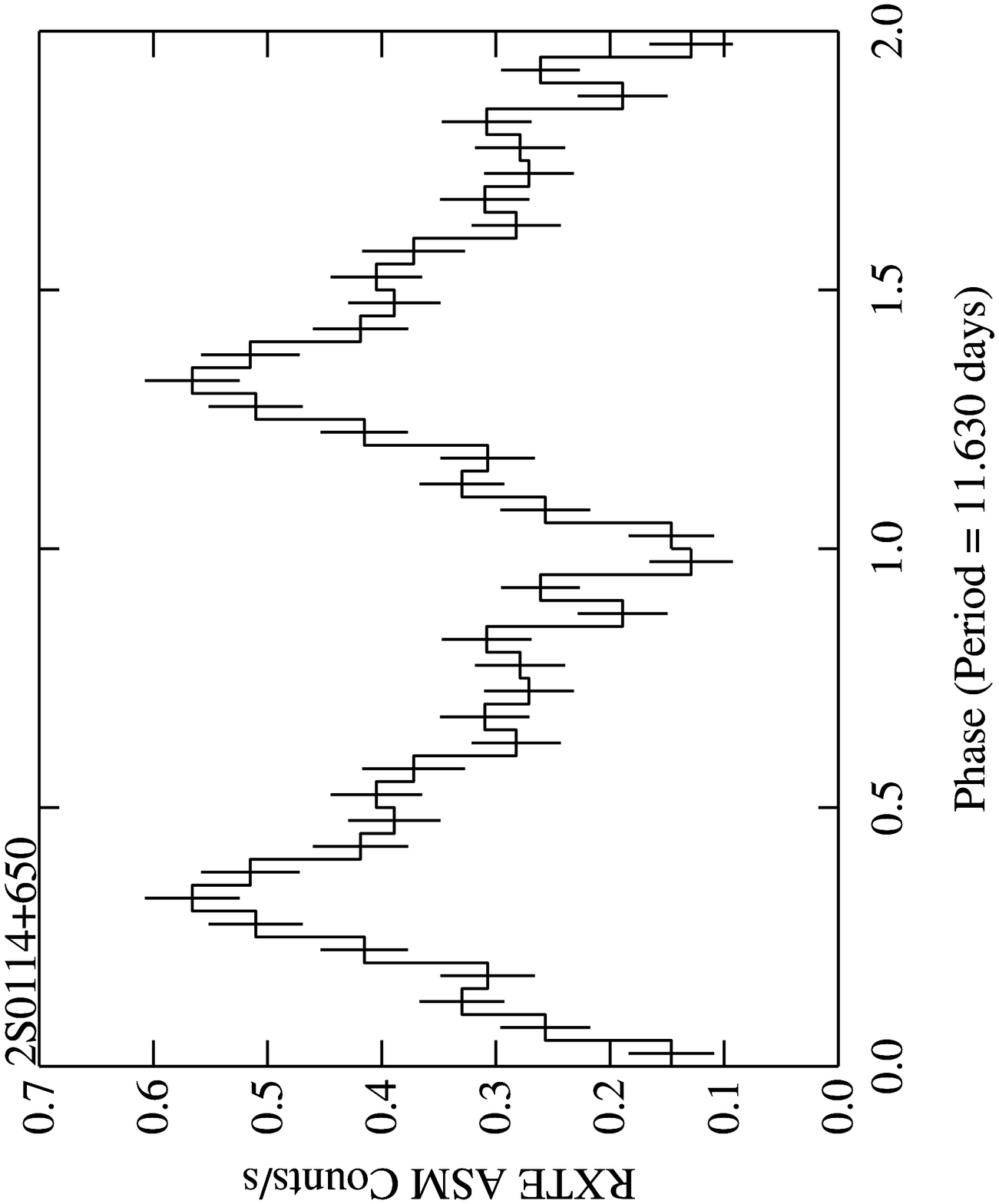]
{ASM light curve folded and binned on the strongest peak in the power
spectrum which is slightly different from the Crampton et al. (1985)
period.}

\figcaption[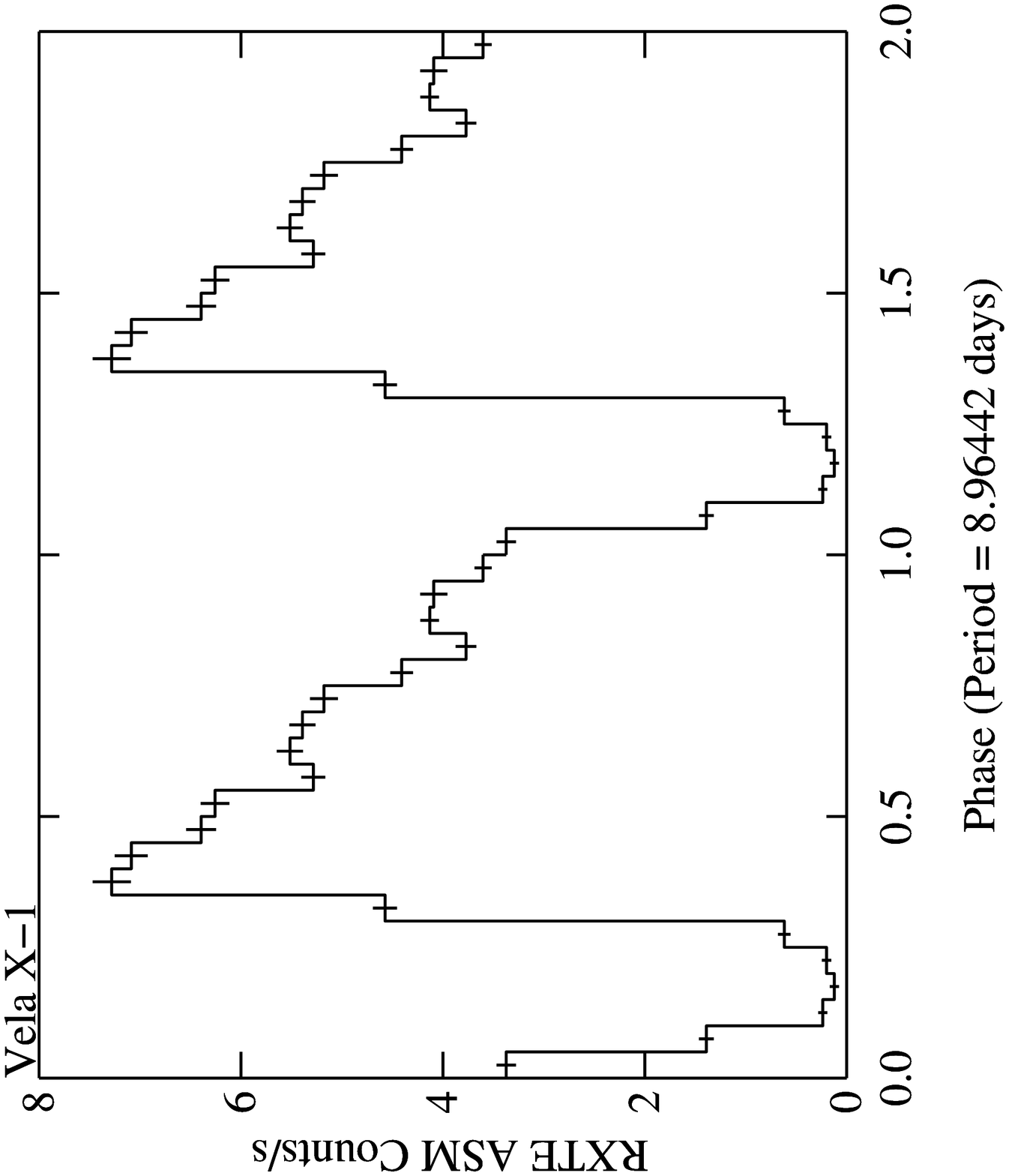]
{ASM light curve of Vela X-1 folded and binned on the orbital period of
this system for comparison with \source. Phase is arbitrary.}


\begin{figure}
\plotone{figure1.ps}
\end{figure}


\begin{figure}
\plotone{figure2.ps}
\end{figure}


\begin{figure}
\plotone{figure3.ps}

\end{figure}


\begin{figure}
\plotone{figure4.ps}
\end{figure}


\begin{figure}
\plotone{figure5.ps}
\end{figure}


\begin{figure}
\plotone{figure6.ps}
\end{figure}

\begin{figure}
\plotone{figure7.ps}

\end{figure}


\begin{figure}
\plotone{figure8.ps}

\end{figure}

\begin{figure}
\plotone{figure9.ps}

\end{figure}

\end{document}